# Importance of Sea Contribution to Nucleons

M. Batra[a], A. Upadhyay[a]

[a]School of Physics and Material Science, Thapar University, Patiala, Punjab -147004

**Abstract**

We studied the statistical model of nucleons consisting of sea having various quark-gluon Fock states in addition to valence quarks. Using statistical consideration and taking 86% of the total Fock states contributing to the low energy properties of nucleon, we aim to find the contributions to these properties coming from the scalar, vector and tensor sea. We checked its validity against the assumption where the contributions from scalar and tensor sea have been suppressed and justified to be unimportant. We took the approximation that sea is getting a zero contribution from $H_0 G_{\overline{10}}$ and $H_1 G_{\overline{10}}$ in three gluon states. Under above considerations, the calculated magnetic moment, spin distribution and weak decay coupling constant ratio for proton and neutron states have been tabulated. We hereby confirm that the suppression of the scalar and tensor sea leads to modification in the parameters of the nucleons showing deviation from the experimental data.

**PACS No**: 12.39. Jh., 12.38 Ee,13.30.Ce.,13.40.Em.,14.20.c

**Keywords**: Nucleon, Phenomenological model, Spin distribution

## 1. Introduction

Hadronic studies are accomplished by a wide range of experiments which produce more précised results about the hadronic structure. Recent experiments at CERN, RHIC, HERMES and COMPASS Collaborations [1-3] measured the spin structure of proton and deuteron system which elaborates the concept of missing spin inside the nucleon. The distinction between quark, anti-quark and gluons contributions to the spin of nucleon is the main puzzle for the recent experiments which lead the phenomenologist to formulate more authentic models which can provide the detailed information about the structure of baryons. The first experimental investigation contributed to structure of baryons was Deep-Inelastic Experiment in which using polarized beams the quark spin distribution of baryons can be probed at typical energies. EMC (European Muon Collaboration) via deep inelastic scattering [4] at CERN concluded that the spin of proton is carried by spin of light quarks and anti-quarks. Later on SMC Collaboration [5] measured the spin structure function $g_1(x)$ of proton and neutron at $Q^2=10$ GeV$^2$. Further a series of experiments at SLAC (for example E-143 Collaboration) [6] measured spin content by valence part and sea-part separately. Some more recent experiments like Drell-Yan processes [7] focus on flavor and spin asymmetries which restrict the symmetry between u and d quarks inside a proton. Recent feedback from different experiments inspires the phenomenologist to improve existing theories so as to go into deeper details of the problem. At very low energy due to inapplicability of perturbative QCD, information regarding the hadronic structure is relatively low. However lattice QCD [8] [9] has also been proved to be useful to gain vast information about low energy data for hadrons. Phenomenologist also tries to propose different models in various ways to satisfy the experimental data and in result, a vast range of models are in hand. For hadrons at low energy, the structure of hadrons in form of quarks was first put forward by Gellman and Zweing [10] and stated that hadrons known at that time could be built up of as a composite system of three quarks (u, d, s) and with fractional charge so as to obey SU(3) symmetry. Later on the picture of colored quarks was formulated by Gell-Mann [11] and observed hadrons are singlet of SU(3) group but was unable to explain the results observed at experiments. The first historical step in understanding of quark-gluon nature was in the constituent quark model[12]. Later on SU(3) states got combined with the SU(2) states of spin and led to six fold symmetry known as SU(6). The model uses six quark states to constitute and classify the hadrons into mesons and baryons and known as SU(6) quark model. SU(6) model explained some of the low energy properties of nucleonic system like magnetic moment ratio still for some parameters for which extension of SU(6) Quark model is needed. Isgur and Karl Model [13] suggested the interaction between quarks in terms of Harmonic oscillator combined with anharmonic perturbation and hyperfine interactions. Bag Model [14] considers the region of space with hadronic fields and having constant energy per unit volume. This region of space is assumed as a "Bag". In the chiral quark model [15] quarks move along with $q\bar{q}$ condensates and an octet of Goldstone bosons is said to be generated as a result of spontaneous broken global symmetry.

One more approach is to visualize the hadronic structure as fundamental quarks as valence part interacting through gluons and quark-antiquark pairs as the active participant in the composition of hadrons. In this paper we focus on the two such models and look for the low energy parameters using two approaches. The two models discussed here assume the protonic structure to be made up of two parts one is valence part and other is sea-part where sea has the structure of quark-antiquark pair multi-connected through gluons. Our models here look the three core quarks in proton state as embedded by the quark-antiquark pair through gluons which is referred here as sea-stuff. Sea-part is assumed to be in S-wave and having non-relativistic nature. The orbital angular momentum between sea part and valence part is considered as negligible here. Two techniques require framing of a suitable wave-function which encloses spin, flavor and color for the valence and sea-part in such a way to give total anti symmetric wave-function which is described in section 2. In section 3, we describe various low energy parameters and their dependence on various coefficients mentioned in the wave-function. Section 4 describes the detailed formalism of the two models, one is statistical model [25] and other is based on simple quark model [27] in which low energy properties are calculated, compared and analyzed. A detailed analysis shows that inclusion of scalar and tensor sea is necessary so as to produce more accurate results.

## 2. Wave Function for Baryon

The total flavor-spin-color wave function of a spin up baryon which consist of three-valence quarks and a sea components can be written as

$$\left|\Phi_{\frac{1}{2}}^{\uparrow}\right\rangle = \frac{1}{N}[\Phi_1^{(\frac{1}{2})^{\uparrow}} H_0 G_1 + a_8 \Phi_8^{(\frac{1}{2})^{\uparrow}} H_0 G_8 + a_{10} \Phi_{10}^{(\frac{1}{2})^{\uparrow}} H_0 G_{\overline{10}} + b_1 \left[\Phi_1^{\frac{1}{2}} \otimes H_1\right]^{\uparrow} G_1 +$$

$$+ b_8 \left(\Phi_8^{\frac{1}{2}} \otimes H_1\right)^{\uparrow} G_8 + b_{10} \left(\Phi_{10}^{\frac{1}{2}} \otimes H_1\right)^{\uparrow} G_{\overline{10}} + c_8 (\Phi_8^{\frac{3}{2}} \otimes H_1)^{\uparrow} G_8 + d_8 (\Phi_8^{\frac{3}{2}} \otimes H_2)^{\uparrow} G_8]$$

Where

$$N^2 = 1 + a_8^2 + a_{10}^2 + b_1^2 + b_8^2 + b_{10}^2 + c_8^2 + d_8^2$$

Any operator $\widehat{O} = \sum_i \widehat{O}_f^i \hat{\sigma}_z^i$ where $\widehat{O}_f^i$ depends upon on the flavor of ith quark and $\hat{\sigma}_z^i$ is the spin projection operator of i$^{th}$ quark where $<\widehat{O}_f^i>^{\lambda\lambda}=\langle\phi^\lambda|O_f^i|\phi^\lambda\rangle$, $<\hat{\sigma}_z^i>^{\lambda\uparrow\lambda\uparrow}=\langle\chi^{\lambda\uparrow}|\sigma_z^i|\chi^{\lambda\uparrow}\rangle$, $<\widehat{O}_f^i>^{\rho\rho}=\langle\phi^\rho|O_f^i|\phi^\rho\rangle$ and $<\hat{\sigma}_z^i>^{\rho\uparrow\rho\uparrow}=\langle\chi^{\rho\uparrow}|\sigma_z^i|\chi^{\rho\uparrow}\rangle$ , $<\widehat{O}_f^i>^{\lambda\rho}=\langle\phi^\lambda|O_f^i|\phi^\rho\rangle$, $<\hat{\sigma}_z^i>^{\lambda\uparrow\rho\uparrow}=$

$\langle \chi^{\lambda\uparrow} | \sigma_z^i | \chi^{\rho\uparrow} \rangle$ etc. where $\lambda$ denotes the symmetric wave-function and $\rho$ denotes the antisymmetry of the wave-function. The wave-function mentioned above can be rewritten in terms of parameters a, b, c, d and e.

$$\left\langle \Phi_{1/2}^{(\uparrow)} \middle| \hat{O} \middle| \Phi_{1/2}^{(\uparrow)} \right\rangle = \frac{1}{N^2}[a \sum_i <\hat{O}_f^i>^{\lambda\lambda} <\hat{\sigma}_z^i>^{\lambda\uparrow\lambda\uparrow} + <\hat{O}_f^i>^{\rho\rho}<\hat{\sigma}_z^i>^{\rho\uparrow\rho\uparrow}+ 2<\hat{O}_f^i>^{\lambda\rho}<\hat{\sigma}_z^i>^{\lambda\uparrow\rho\uparrow}+b\sum_i(<\hat{O}_f^i>^{\lambda\lambda}+ <\hat{O}_f^i>^{\rho\rho})(<\hat{\sigma}_z^i>^{\lambda\uparrow\lambda\uparrow}+<\hat{\sigma}_z^i>^{\rho\uparrow\rho\uparrow}) + c \sum_i[<\hat{O}_f^i>^{\lambda\lambda}<\hat{\sigma}_z^i>^{\rho\uparrow\rho\uparrow}+ <\hat{O}_f^i>^{\rho\rho}<\hat{\sigma}_z^i>^{\lambda\uparrow\lambda\uparrow}- 2<\hat{O}_f^i>^{\lambda\rho}<\hat{\sigma}_z^i>^{\rho\uparrow\rho\uparrow}] + d[\sum_i <\hat{O}_f^i>^{\lambda\lambda}+\sum_i <\hat{O}_f^i>^{\rho\rho}] + e[\sum_i (<\hat{O}_f^i>^{\rho\rho}-<\hat{O}_f^i>^{\lambda\lambda}) <\hat{\sigma}_z^i>^{\lambda\uparrow\frac{3}{2}\uparrow}+ 2\sum_i <\hat{O}_f^i>^{\lambda\rho}<\hat{\sigma}_z^i>^{\rho\uparrow\frac{3}{2}\uparrow}]$$

The following coefficients are defined as

$a=\frac{1}{2}(1-\frac{b_1^2}{3})$, $b=\frac{1}{4}(a_8^2-\frac{b_8^2}{3})$, $c=\frac{1}{2}(a_{10}^2-\frac{b_{10}^2}{3})$, $d=\frac{1}{18}(5c_8^2-3d_8^2)$, $e=\frac{\sqrt{2}}{3}b_8c_8$.

The properties like magnetic moment, spin distribution, weak decay coupling constant ratios are calculated by defining suitable operators for flavor and spin part. Suitable expression is obtained from the eigenvalues coming from the above defined operators. For instance, substituting the spin operator in spin ½ proton wave-function when it operates on symmetric part of wave-function gives 2/3 and operating on antisymmetric wave-function gives 0.

$$\langle \chi^{\lambda\uparrow} | \sigma_z^i | \chi^{\lambda\uparrow} \rangle = \left\langle \frac{1}{\sqrt{6}}(\uparrow\downarrow+\downarrow\uparrow)\uparrow -2\uparrow\uparrow\downarrow \middle| \sigma_z^{(1)} \middle| \frac{1}{\sqrt{6}}(\uparrow\downarrow+\downarrow\uparrow)\uparrow -2\uparrow\uparrow\downarrow \right\rangle = \frac{2}{3} \text{ and}$$

$$\langle \chi^{\rho\uparrow} | \sigma_z^i | \chi^{\rho\uparrow} \rangle = \left\langle \frac{1}{\sqrt{2}}(\uparrow\downarrow-\downarrow\uparrow)\uparrow \middle| \sigma_z^{(1)} \middle| \frac{1}{\sqrt{2}}(\uparrow\downarrow-\downarrow\uparrow)\uparrow \right\rangle = 0.$$

The generalized expressions in terms of two parameters $\alpha$ and $\beta$ may be useful for studying the above said low energy properties of a hadronic system.

$$\alpha = \frac{1}{N^2}\left(\frac{4}{9}\right)(2a+2b+3d+\sqrt{2}e)$$

$$\beta = \frac{1}{N^2}\left(\frac{1}{9}\right)(2a-4b-6c-6d+4\sqrt{2}e)$$

Or

$$\alpha = \frac{2(6+3a8^2-2b1^2-b8^2+4b8c8+5c8^2-3d8^2)}{27(1+a10^2+a8^2+b1^2+b10^2+b8^2+c8^2+d8^2)}$$

$$\beta = \frac{3 - 9a10^2 - 3a8^2 - b1^2 + 3b10^2 + b8^2 + 8b8c8 - 5c8^2 + 3d8^2}{27(1 + a10^2 + a8^2 + b1^2 + b10^2 + b8^2 + c8^2 + d8^2)}$$

Thus the two parameters directly relate the various sea contributions in terms of various co-efficient. The physical significance of two parameters lies in their relation with number of spin up and spin down quarks in the spin up baryon that is $\Delta q = n(q \uparrow) - n(q \downarrow) + n(\bar{q} \uparrow) - n(\bar{q} \downarrow)$, $q = u, d, s$. $\Delta q$ is used to find the spin structure of proton, weak decay coupling constant ratios and also there exists a class of models [16] which relate $\Delta q$ with the magnetic moment of baryons. This makes $\alpha$ and $\beta$ as the key parameters to solve following low energy properties of nucleons.

## 3. Nucleonic Parameter at Low Energy

### 3.1 Spin Distribution of a Nucleonic System

The investigation of spin structure of proton is related to the distribution of spin among the valence quark, sea-quark and gluonic constituent inside a proton. Deep inelastic scattering experiments shows that 30% of spin of proton is carried by spin of its quark constituents. Later on SMC (Spin Muon Collaboration) [17] published their measurement of spin dependent structure functions g1 and suggested that quark's intrinsic spin contributes very less to proton spin. The measurement of spin distribution function by EMC experiment in 1988 measured $\Gamma_p^1$ as 0.126±0.018 [18]. Ellis and Jaffe sum rule predicts the same spin distribution function using SU(3) flavor symmetry [19]. As per helicity sum rule, the contributions of spin comes not only from up, down and strange quark spins but gluonic spins and angular momentum of quarks and gluons also.

$$S_Z = \frac{1}{2} = \frac{1}{2}(\Delta u + \Delta d + \Delta s) + \Delta G + L_Z$$

Where the quark distributions are contained in term $\Delta\Sigma = \Delta u + \Delta d + \Delta s$ and $\Delta G$ is the contribution from gluonic spin and $L_Z$ is the orbital angular momentum of quarks and gluons recent experimental studies are governed by COMPASS and HERMES collaborations focus on the measurement of $\Delta\Sigma = 0.35 \pm 0.03(stat) \pm 0.05(sys)$ and $\Delta\Sigma = 0.330 \pm 0.011(theo) \pm 0.025(exp) \pm 0.028(evol)$[20] [21] at $Q^2$ =3 GeV$^2$ and 5 GeV$^2$ respectively. At low $Q^2$, the higher twist corrections also become dominant and affect the value of low energy parameters. E. Leader et al. [22] applied HT corrections to low energy parameters using DIS experimental techniques and results obtained are found to be matching with QCD sum rules and instanton model predictions.

We here obtain $I_1^p = \frac{1}{6}(4\alpha - \beta)$ and $I_1^n = \frac{1}{6}(\alpha - 4\beta)$ as spin distributions in proton and neutron which is obtained using charge squared spin projection operator $I_1^p = \frac{1}{2} < \sum_i e_i^2 \sigma_z^i >_p$ and $I_1^n = \frac{1}{2} < \sum_i e_i^2 \sigma_z^i >_n$. The operator here provides contribution in terms of quark and gluon Fock states and furthermore, the orbital angular momentum share is considered negligible here due to very less overlap regions between momentum of valence and sea [23].

### 3.2 Weak Decay Coupling Constant Ratio

The matrix element of quark currents between proton and neutron states in beta decay can be calculated using isospin symmetry. From the semileptonic decay B→B`e ν, vector and axial coupling constants $g_V$ and $g_A$ can be determined. On the basis of SU(3) symmetry, all baryon decay rates depend upon two universal parameters F and D where the sum F+D is defined as $\lambda = g_A/g_V$. Operator for this ratio can be taken as: $\hat{O}_f^i = 2I_3^i$ and this gives $g_A/g_V = 3(\alpha + \beta)$ for neutron decaying into proton. The ratio of F and D is interpreted as $\frac{F}{D} = \frac{\alpha}{\alpha + 2\beta}$.

### 3.3 Magnetic Moment Ratio

Similarly, the magnetic moment ratio of proton and neutron in terms of $\alpha$ and $\beta$ is $\frac{\mu_P}{\mu_N} = -\frac{2\alpha + \beta}{\alpha + 2\beta}$.

Magnetic moment ratio of proton and neutron is obtained using $\hat{\mu} = \sum_q \frac{e_q^i}{2m} \sigma_z^q, q = (u, d, s)$ giving $\mu_p = 3(\mu_u \alpha - \mu_d \beta)$ and $\mu_n = 3(\mu_d \alpha - \mu_u \beta)$.

Thus nucleonic properties at low energy are expressed in terms of above defined parameters in the following manner:

$$I_1^p = \frac{(15 + 3a_{10}^2 + 9a_8^2 - 5b_1^2 - b_{10}^2 - 3b_8^2 + 8b_8c_8 + 15c_8^2 - 9d_8^2)}{54(1 + a_{10}^2 + a_8^2 + b_1^2 + b_{10}^2 + b_8^2 + c_8^2 + d_8^2)}$$

$$I_1^n = \frac{(3a_{10}^2 + 9a_8^2 - 5b_1^2 - b_{10}^2 - 3b_8^2 + 8b_8c_8 + 15c_8^2 - 9d_8^2)}{54(1 + a_{10}^2 + a_8^2 + b_1^2 + b_{10}^2 + b_8^2 + c_8^2 + d_8^2)}$$

$$\frac{g_A}{g_V} = \frac{(15 + 3a_8^2 - 9a_{10}^2 - 5b_1^2 + 3b_{10}^2 - b_8^2 + 16b_8c_8 + 5c_8^2 - 3d_8^2)}{9(1 + a_{10}^2 + a_8^2 + b_1^2 + b_{10}^2 + b_8^2 + c_8^2 + d_8^2)}$$

$$\frac{F}{D} = \frac{(21a_8^2 + 9a_{10}^2 + 25b_1^2 + 33b_{10}^2 + 29b_8^2 + 16b_8c_8 + 17c_8^2 + 33d_8^2 + 33)}{27(1 + a_{10}^2 + a_8^2 + b_1^2 + b_{10}^2 + b_8^2 + c_8^2 + d_8^2)}$$

$$\frac{\mu_p}{\mu_n} = \frac{(9 + 3a_8^2 - 3a_{10}^2 - 3b_1^2 + b_{10}^2 - b_8^2 + 8b_8c_8 + 5c_8^2 - 3d_8^2)}{(6 - 6a_{10}^2 - 2b_1^2 + 2b_{10}^2 + 8b_8c_8)}$$

An elaborated analysis of low energy parameters shows that some of the coefficients in baryonic wavefunction have a zero dependence for α or β. Some of the terms like $a_{10}$ and $b_{10}$ is missing in the parameter α and F/D ratio but $b_8$ and $c_8$ term always appear with maximum contribution in the form of $b_8c_8$, which signifies the maximum contribution from gluon with spin 1. The co-efficient $b_1$ seems to be non-contributing to spin distribution of neutron. Contribution to all low energy parameters from tensor sea comes from only one term that is $d_8$ and it always appears with negative sign (except $I_1^n$).

## 4. Phenomenological Models

There exists a long list of models (non-relativistic as well as relativistic) which calculate these low energy parameters through non-perturbative sea quark components in baryons. These models either describe the hadrons in terms of fundamental quarks as valence part and quark-antiquark pair along with gluons as the active participant in the composition of baryons or some models considers virtual meson-baryon states as the contribution to the nucleon structures. Our main concern here is the comparative study of two different models assuming sea as made up of admixture of gluons and quark-antiquark pair.

The one of the two approaches discussed here includes statistical model which uses principle of balance and detailed balance [23][24] which succeeds in explaining the more recent phenomenon that is $\bar{u}$ and $\bar{d}$ asymmetry. This principle calculates $\bar{d}$- $\bar{u}$ =0.124 which lies in close agreement with the experimental data value that is 0.118±0.012 [29]. It considers proton to be made up of different Fock states and the probability associated with each Fock states is further utilized to compute the coefficients which contribute to low energy properties but the second method does not focus on calculating such kind of probabilities rather estimate the coefficients by just fitting the low energy parameters.

The statistical model views proton state as a complete set of quark-gluon Fock states as:-

$$|p\rangle = \sum_{i,j,k} C_{ijk} |uud, i, j, k\rangle$$

Where $i$ is the number of quark-anti-quark $u\bar{u}$ pair, $j$ is the number of $d\bar{d}$ pairs and $k$ is the number of gluons. The probability of finding the proton in Fock state $|uud, i, j, k\rangle$ is:

$$\rho_{i,j,k} = |C_{i,j,k}|^2$$

The methodology consists of finding the relative probability for $q^3$ core and sea-part where sea-part can have different possibilities. The relative probability in spin and color space is taken for each of the case so as the core part should have an angular momentum as $j_1$ and sea-content $j_2$ and the total angular momentum should be $j_{1+}\, j_{2=}½$ and resultant should be in a color singlet state. These probability ratios are expressed in the form of a common parameter c. "c" has its own significance in the sense that parameters $\alpha$ and $\beta$ are computed using the value of "nc" where n represents the respective multiplicities for different gluonic states. The contributions from $H_0 G_{\overline{10}}$ and $H_1 G_1$ sea for two gluon states is not considered as $H_0$ and $G_1$ are symmetric whereas $H_1$ and $G_{\overline{10}}$ are anti-symmetric under the exchange of two gluons which makes these product wave-function anti-symmetric which is unacceptable for the bosonic system. A single gluon in the sea will contribute only to the $H_1 G_8$ component of the sea while other states cannot include this due to symmetry conditions [26]. We here apply some modifications to the previous calculations in the sense that the contributions from three gluon states is taken as zero for $H_0 G_{\overline{10}}$ and $H_1 G_{\overline{10}}$ as well as for the states which assumes the spin of valence part as 3/2. Thus the modified value of the parameters give value $\alpha = 0.217$ and $\beta = 0.0715$. From α and β, all the low energy nucleonic parameters can be calculated.

On the other hand, simple quark model which was originally proposed by Li [27] and extended by Song and Gupta [28] computes $\alpha$ and $\beta$ analytically using available data as input. The magnetic moments of all the baryons can be expressed in terms of quark magnetic moment and the coefficients $\alpha$ and $\beta$ in following ways: $\mu_p = 3(\mu_u \alpha - \mu_d \beta)$ and $\mu_n = 3(\mu_d \alpha - \mu_u \beta)$ and so on. Using ($\widetilde{U} = 3\alpha\, \mu_d$, $\widetilde{D} = -3\beta\mu_d$, $2p = -\mu_u/\mu_d$, $r = \mu_s/\mu_d$), we express baryonic magnetic moments in terms of these four parameters. We predict $\frac{\alpha}{\beta} = 4.406$ and $\frac{F}{D} = \frac{\alpha}{\alpha+2\beta} = \frac{\frac{\alpha}{\beta}}{\frac{\alpha}{\beta}+2} = 0.688$. From this, $\alpha$ and $\beta$ can be separated out which gives $\alpha = 0.3415$ and $\beta = 0.077$. With this technique, the above calculated parameters are used to estimate the contributions from scalar, vector and tensor sea. The difference between the two methods is that vector sea is considered as only active contributor in sea-part in simple quark model. Motivation to neglect tensor sea comes from the statement that tensor sea contribution comes from spin 3/2 valence part and it becomes less probable for core part to have spin 3/2. Song and Gupta assumed that if only scalar sea contributes then the results do not match with the experimental results. Vector sea dominance comes in the form of just four fitted parameters as: $b1 = 0.0642, b8 = 0.47, c8 = 0.16, a10 = 0.31$.

## 5. Discussion of Results

A table showing results from two models are given below:

| Low energy parameters | Statistical Model[25] | Song and Gupta[27] with zero sea | Statistical model keeping scalar and tensor part as zero | Song and Gupta[27] keeping scalar and tensor part as non-zero | Experimental Result with Errors |
|---|---|---|---|---|---|
| $\alpha$ | 0.217 | 0.3415 | 0.2569 | 0.225 | ..... |
| $\beta$ | 0.07145 | 0.07749 | 0.0913 | 0.041734 | ...... |
| $\mu_p/\mu_n$ | -1.41 | -1.5317 | -1.37 | -1.59 | -1.46[30] |
| $g_A/g_V$ | 0.8655 | 1.26 | 1.041 | 0.803 | 1.257±0.03[33] |
| F/D | 0.603 | 0.6878 | 0.5846 | 0.7298 | 0.575 ± 0.016 [33] |
| $I_1^P$ | 0.132 | 0.214726 | 0.156 | 0.14336 | 0.132±0.003±0.009 [32] |
| $I_1^n$ | -0.01145 | 0.0052452 | -0.00181 | 0.009756 | -0.030 |

Table 5.1-Results from different approaches for low energy parameters are shown here.

Statistical model calculations come here with inclusion of Fock states upto 86%. Rest part may include $s\bar{s}$ content and higher Fock states have been neglected that leads to a negligible contribution.. Further, comparison from the results given by Song and Gupta [28] as well as experimental data is shown in table 5.1. Along with this, the effect of suppressing the contribution from scalar and vector sea in statistical model and vice-versa has also been analyzed. Still one cannot think of ignoring the parameters completely in calculation of all the properties of a baryonic system. Each term has non negligible dependence on almost all the nucleonic parameters. On neglecting the scalar and tensor sea completely, all the properties are not satisfying the experimental results simultaneously, even inclusion of these contributions does not completely justify the results. For instance, neglecting the scalar and tensor sea in statistical model calculations deviate the results by ≅3-10% as compared to original results in all cases except weak decay matrix element ratio, that deviates 3.5% from original value. On the other hand, if we take into account the scalar and tensor sea in Simple quark model then the percentage error from the experimental results increase upto 7-8% but drastic change is observed in spin distribution to the nucleons. Here the deviation goes down from ~50 % to 6-7 % approximately.

Moreover, a graphical analysis of low energy properties with respect to different parameters like $a_8$, $a_{10}$, $b_1$, $b_8$, $b_{10}$ is shown below. Here to check the contribution from the scalar sea, we suppress the vector and tensor sea contributions and similar approach to find the individual contribution from vector and tensor sea.

1. The plots with respect to the parameter $a_8$, $a_{10}$ predict that values of α, β and other parameters decrease to a large extent but a sufficient variation is observed in weak decay coupling constant ratio.

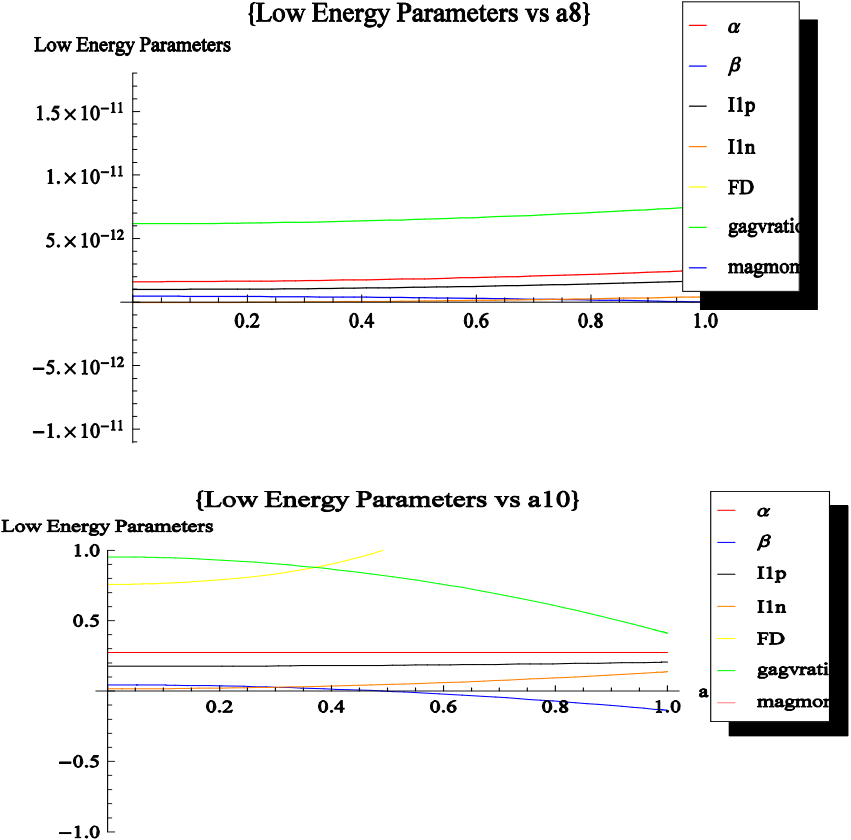

Fig. 5.1- Low energy parameters for nucleons Vs coefficients contributing to scalar sea

2. The graphical interpretation here shows that maximum variation of parameters can be seen due to vector sea. As the sea part is dominated by emission of virtual gluons so we can expect $b_8$ and $c_8$ to be more varying in nature. If only vector sea is assumed to be dominating, nucleonic properties like coupling constant and F/D ratio are mainly affected with parameters $b_8$ and $c_8$.

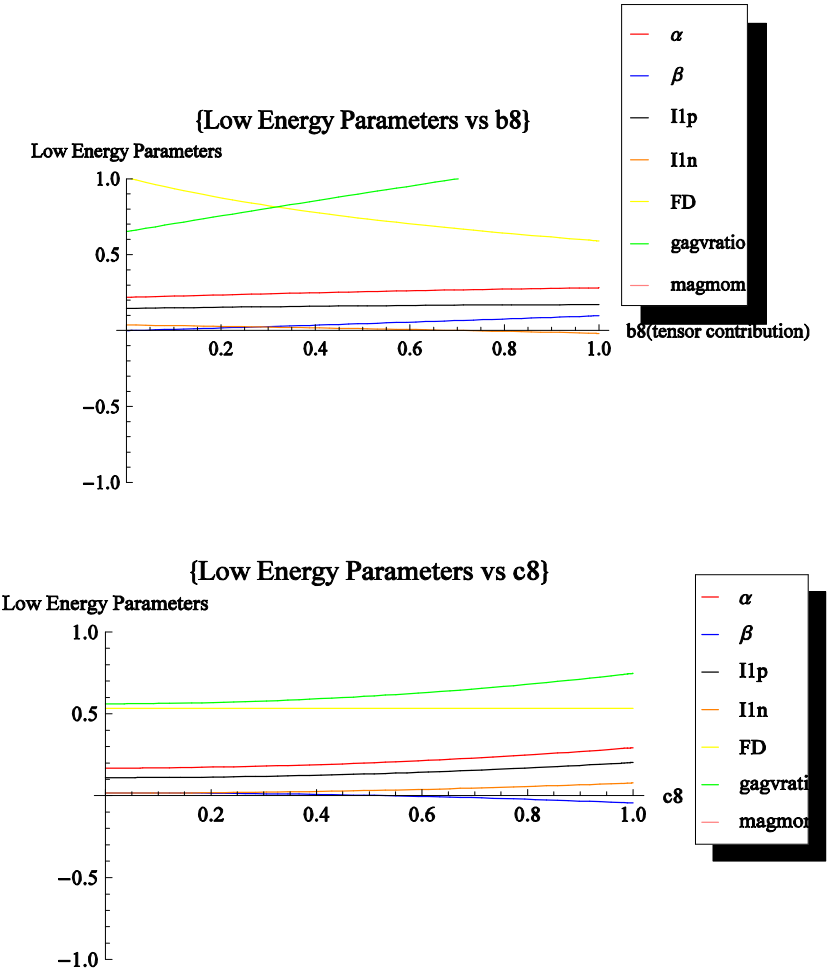

Fig. 5.2- Low energy parameters for nucleons Vs coefficients contributing to vector sea.

3. Tensor sea appears to be less dominating due to quark-spin flip process but cannot be neglected in all cases. Some of the properties seem to be the most affected by the change in the values of these coefficients.

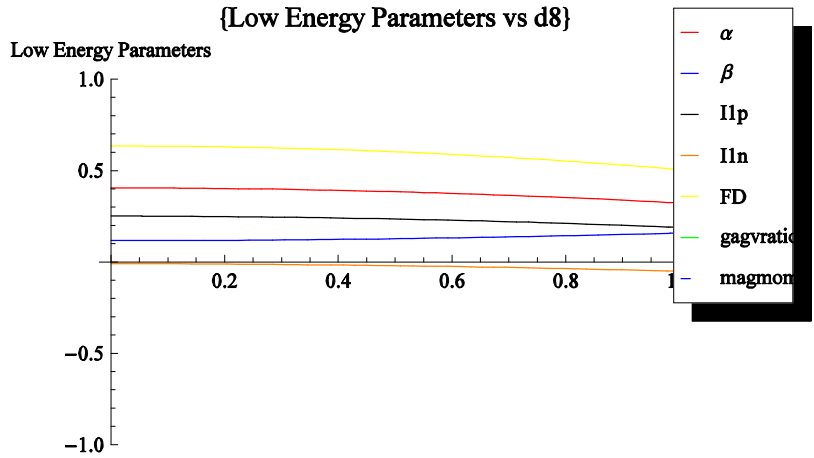

Fig. 5.3- Low energy parameters for nucleons Vs coefficients contributing to tensor sea.

Our calculation holds good for low energy scale of the order of 1GeV$^2$ but experimental results are applied at typical energy scale Q$^2$~10 GeV$^2$. Hence the deviation are within 10% of actual value however large deviations in the value of spin distribution may be due to reason that our calculations are performed in baryonic rest frame and angular momentum corrections may produce more accurate results. We are also aiming to calculate the similar properties for other baryonic octet by taking the contribution also from the strange $q\bar{q}$ condensates present in the sea.